\documentclass[twocolumn,prl,aps,superscriptaddress]{revtex4}

\usepackage{amsmath,amssymb,latexsym,revsymb}

\def\duzomniejsze{<\kern-.7mm<}
\def\duzowieksze{>\kern-.7mm>}

\def\textbf#1{{\bf #1}}
\def\beq{\begin{equation}}
\def\eeq{\end{equation}}
\def\be{\begin{equation}}
\def\ee{\end{equation}}
\def\ben{\begin{eqnarray}}
\def\een{\end{eqnarray}}
\def\beqa{\begin{eqnarray}}
\def\eeqa{\end{eqnarray}}
\def\eea{\end{array}}
\def\bea{\begin{array}}
\newcommand{\bei}{\begin{itemize}}
\newcommand{\eei}{\end{itemize}}
\newcommand{\bee}{\begin{enumerate}}
\newcommand{\eee}{\end{enumerate}}

\newcommand{\tr}{\operatorname{Tr}}
\newcommand{\ID}{\operatorname{id}}

\def\>{\rangle}
\def\<{\langle}

\def\ot{\otimes}

\def\T{Z}
\def\R{Z}
\def\Z{Z}
\def\px{P_X}

\def\pxyz{P_{XYZ}}
\def\co{{\cal C}_o}
\def\ci{{\cal C}_i}
\newcommand{\ket}[1]{| #1 \rangle}
\newcommand{\bra}[1]{\langle #1 |}
\newcommand{\proj}[1]{\ket{#1}\bra{#1}}

\newtheorem{lemma}{Lemma}

\newtheorem{theorem}[lemma]{Theorem}
\newtheorem{definition}[lemma]{Definition}

\begin{document}

\title{A classical analogue of negative information}

\begin{abstract}
  Recently, it was discovered that the {\it quantum partial information}
  needed to merge one party's state with another party's state
  is given by the conditional entropy, which can be negative
  [Horodecki, Oppenheim, and Winter, Nature {\bf 436}, 673 (2005)].
  Here we find a classical analogue of this, based on a long known relationship
  between entanglement and shared private correlations: namely,
  we consider a private distribution held between two parties, and
  correlated to a reference system, and ask how much secret communication
  is needed for one party to send her distribution to the other.  We give
  optimal protocols for this task, and find that private information can be
  negative -- the sender's distribution can be transferred and the potential
  to send future distributions in secret is gained through the
  distillation of a secret key.  An analogue of {\it quantum state exchange}
  is also discussed and one finds cases where exchanging a distribution
  costs less than for one party to send it. The results give new classical
  protocols, and also clarify the various relationships between
  entanglement and privacy.
\end{abstract}

\author{Jonathan Oppenheim}
\affiliation{Department of Applied Mathematics and Theoretical Physics, University of Cambridge U.K.}

\author{Robert W. Spekkens}
\affiliation{Perimeter Institute for Theoretical Physics,
             31 Caroline St. N, Waterloo, Ontario N2L 2Y5, Canada}

\author{Andreas Winter}
\affiliation{Department of Mathematics, University of Bristol, Bristol BS8 1TW, U.K.}

\date{28 November 2005}

\maketitle

{\bf Introduction.} While evaluating the quality of information is
difficult, we can quantify it.  This was first done by Shannon
\cite{Shannon1948} who showed that the amount of information of a
random variable $X$ is given by the Shannon entropy $H(X)=-\sum
\px(x) \log_2 \px(x)$ where $\px(x)$ is the probability that the
source produces $X=x$ from distribution $\px$. If $n$ is the
length of the message (of independent samples of $X$) we want to
communicate to a friend, then $\sim nH(X)$ is the number of bits
required to send them.  If our friend already has some prior
information about the message we are going to send him (in the
form of the random variable $Y$), then the number of bits we need
to send him is less, and is given by $n$ times the conditional
entropy $H(X|Y)=H(XY)-H(Y)$, according to the Slepian-Wolf theorem
\cite{slepian-wolf}.

In the case of quantum information, it was shown by Schumacher
\cite{Schumacher1995} that for a source producing a string of $n$ unknown
quantum states with density matrix $\rho_A$, $\sim nS(A)$ quantum
bits (qubits) are necessary and sufficient to send the states
where $S(A)=-\tr\rho_A\log\rho_A$ is the von Neumann entropy (we
drop the explicit dependence on $\rho$ in $S(A)$).  One can now
ask how many qubits are needed to send the states if the receiver
has some prior information.  More precisely, if two parties, Alice
and Bob, possess shares $A$ and $B$ of a bipartite system $AB$
described by the quantum state $\rho_{AB}$, how many qubits does
Alice need to send Bob so that he can locally prepare a bipartite
system $A'B$ described by the same quantum state (classical
communication is free in this model). We say that Bob has some
prior information in the form of state $\rho_B=\tr\rho_{AB}$, and
Alice wants to {\it merge} her state with his by sending him some
{\it partial quantum information}.

Recently, it was found that a rate of $S(A|B)=S(AB)-S(B)$ qubits
are necessary and sufficient \cite{how-merge,how-merge2} for this
task. More mathematically: just as in Schumacher's quantum source
coding \cite{Schumacher1995}, we consider a source emitting a sequence of
$n$ unknown states, but the statistics of the source, i.e. the
average density matrix of the states, is known. The ensemble of
states which realize the density matrix is however unspecified. We
then demand that the protocol allows Alice to transfer her share
of the state to Bob with high probability for all possible states
from the ensemble. A more compact way to say this is to imagine
that the state which Alice and Bob share is part of some pure
state shared with a reference system $R$ and given by
$\ket\psi_{ABR}$ such that $\rho_{AB}$ is obtained by tracing over
the reference system. A successful protocol will result in
$\rho_{AB}^{\otimes n}$ being with Bob, and
$\ket\psi_{ABR}^{\otimes n}$ should be virtually unchanged, while
entanglement is consumed by the protocol at rate $S(A|B)$.

The quantity, $S(A|B)$ is the quantum conditional entropy, and it
can be negative \cite{Wehrl78,HH94-redun,cerfadami}. This
seemingly odd fact now has a natural interpretation \cite{how-merge} -- the
conditional entropy quantifies how many qubits need to be sent
from Alice to Bob, and if it is negative, they gain the
potential to send qubits in the future at no cost. That is, Alice
can not only send her state to Bob, but the parties are
additionally left with maximally entangled states which can be
later be used in a teleportation protocol to transmit quantum
states without the use of a quantum channel. This is the operational
meaning of the fact that partial information can be negative in
the quantum world.

\medskip
{\bf A classical model.} In order to further understand the notion
of negative information, we are interested in finding some
classical analogue of it.  Indeed we will find a paradigm in which
not only is there a notion of negative information, but also the
rate formulas and proof techniques are remarkably similar. We
shall take as our starting point the similarity between
entanglement and private correlations, a fact that was used in
constructing the first entanglement distillation protocols, was used
to conjecture new types of classical distributions \cite{gisin-wolf-99}, but
which was first made fully explicit by Collins and Popescu
\cite{collins-popescu}. In this paradigm, maximally entangled
states are replaced by perfect secret correlations (a ``key'')
$\overline\Psi$, with probability distribution
${\overline\Psi}_{XY}(0,0)={\overline\Psi}_{XY}(1,1)=\frac{1}{2}$.
By \emph{secret}, we mean that a third party, an eavesdropper Eve,
is uncorrelated with Alice and Bob's secret bit.  We then replace
the notion of classical communication by public communication
(i.e., the eavesdropper gets a copy of the public messages that
Alice and Bob send to each other). Quantum communication (the
sending of coherent quantum states) is replaced by secret
communication, i.e. communication through a secure channel such
that the eavesdropper learns nothing about what is sent.  We thus
have sets of states (i.e. classical distributions between various
parties and an eavesdropper), and a class of operations -- local
operations and public communication (LOPC).  Under LOPC one cannot
increase secrecy, just as under local operations and classical
communication (LOCC) one cannot increase entanglement.
The analogy has the essential feature, as in entanglement theory,
that there is a resource (secret key, pure entanglement)
which allows for the transfer of information (private distributions, quantum states),
and this information can be manipulated (by means of classical or public information),
and transformed into the resource.
This allows for the possibility of negative information.
We will further be able to make new statements about
the analogy. For example, we will find indications for 
an analogue of pure states, mixed states,
and various types of GHZ states \cite{ghz}.

Looking at the quantum model, we should consider an arbitrary
distributed source between Alice and Bob, described by a pair of
random variables with probability distribution $P_{XY}$;
furthermore we need a ``purification'', that is an extension of
this distribution to a distribution $P_{XYZ}$ with $Z$ being held
by a party $R$, which we call the reference (who has the marginal
distribution $P_Z$). According to this and
\cite{collins-popescu}, the natural approach will be as follows.  A
pure quantum state held between two parties has a Schmidt
decomposition
$\ket\psi_{TR}=\sum_i\sqrt{p(i)}\ket{e_i}\ot\ket{f_i}$, with
orthonormal bases $\{ \ket{e_i} \}$ and $\{ \ket{f_i} \}$. An
analogue of this is a private \emph{bi-disjoint distribution}, i.e. a
distribution $P_{T\Z}$ (where $T \equiv XY$),
\beq
  \label{eq:bi-disjoint}
  P_{T\Z}(tz)=\sum_i p(i)\,P_{T|I=i}(t)P_{Z|I=i}(z),
\eeq
with conditional distributions $P_{Z|I}$ and $P_{T|I}$, such
that $P_{T|I=i}(t)P_{T|I=j}(t)=0$ and $P_{Z|I=i}(z)P_{Z|I=j}(z)=0$
for $i\neq j$. Just as the quantum system $TR$ is in a product
state between $T$ and $R$ once $i$ is known, so the bi-disjoint
distribution is in product form
$P_{TZ|I=i}(tz)=P_{T|I=i}(t)P_{Z|I=i}(z)$ once $i$ is known. And
just as a pure quantum state is decoupled from any environment, so
our distribution should be decoupled from the eavesdropper. Note
that it appears necessary here to introduce a fourth party $E$,
something we could avoid in the quantum setting by demanding that
the overall pure state is preserved -- for distributions the
meaning of this is staying decoupled from the eavesdropper, which
we have to distinguish from the reference \cite{foot-1}.
Introducing the eavesdropper into the notation, we have
$P_{XYZE}=P_{XYZ} \otimes P_E$. Such distributions we call
\emph{private}, meaning that $E$ is decoupled. In that regard, we
shall speak of \emph{secret} distributions (between Alice and Bob)
where they are decoupled from $R$ and $E$ -- following terminology
introduced on~\cite{csiszar:narayan}. We will provide further
justification for the appropriateness of this analogue of pure
states after we have fully analysed merging and negative
information \cite{foot-2}.  Note however, that it has the following desired property:
in the quantum case, considering a purification of the $AB$ system allows
us to enforce the requirement that the protocol succeed for particular pure
state decompositions of $\rho_{AB}$.  Likewise the distribution $P_{XYZ}$
allows us to enforce the requirement that the protocol succeed for a decomposition
of the distribution $P_{XY}$, with the record being held by $R$.

We now introduce the analogue of quantum state merging -- {\it
distribution merging} -- which naturally means that at the end Bob
and the reference should possess a sample $\widehat{X}\widehat{Y}Z$ from
the distribution $P_{XYZ}$, with $Z$ held by the reference and
$\widehat{X}\widehat{Y}$ by Bob. The protocol may use public communication
freely; we will consider only the rate of secret key used or
created. We also go to many copies of the random variables -- thus
we denote by $X^n$ many independent copies of random variable
$X$, while $\widehat{X}^n$ denotes the output sample of length $n$.
Formally:
\begin{definition}
  Given $n$ instances of a private bi-disjoint distribution
  $P_{XYZ}$ between $AB$ and $R$,
  a {\it distribution merging} protocol between a sender who holds $X$
  and receiver who holds $Y$, is one which creates, by possibly
  using $k$ secret key bits and free public communication, a
  distribution $P_{\widehat{U}^l(\widehat{X}^n\widehat{Y}^n\widehat{V}^l)Z^n\widehat{E}^n}'$
  such that $P'$ approximates
  $P_{XYZE}^{\otimes n} \otimes {\overline\Psi}_{UV}^{\otimes l}$
  for large $n$ (in total variational, or $\ell^1$, distance).
  Here $l$ is the number of secret bits shared at the end between
  Alice and Bob; Alice has $\widehat{U}^l$ and Bob $\widehat{V}^l\widehat{X}^n\widehat{Y}^n$.
  \par\noindent
The rate of consumption of secret key for the protocol, called its
\emph{secret key rate}, is defined to be $\frac{1}{n}(k-l)$.
\end{definition}
We can now state our main result:
\begin{theorem}
  \label{thm:main}
  A secret key rate of
  \beq
    \label{eq:rate}
    I(X:Z)-I(X:Y) = H(X|Y)-H(X|Z)
  \eeq
  bits is necessary and sufficient to achieve distribution merging.
  Here, $I(X:Y):= H(X)+H(Y)-H(XY)$
  is the mutual information.
  When this quantity is nonnegative, it is the minimum rate of secret
  key consumed by an optimal merging protocol.
  When it is negative, not only is distribution merging
  achieved, but $I(X:Y)-I(X:Z)$
  bits of secret key remain at the end of the protocol.
\end{theorem}

Before proving this theorem, and
introducing the protocol in full generality, it may be useful to
discuss three very simple examples:
\begin{enumerate}
  \item Alice's bit is independent of Bob's bit, but correlated with Eve:
    $P_{XYZ}(0,0,0) \!=\! P_{XYZ}(1,0,1) \!=\! \frac{1}{2}$
    In this case, Alice must send her bit to Bob through a secret channel, consuming
    one bit of secret key.

  \item Alice and Bob have a perfect bit of shared secret correlation:
    Bob can locally create a random
    pair of correlated bits, and Alice and Bob keep the bit
    of secret correlation as secret key (which they may use in the
    future for private communication).  There is one bit of negative information.

  \item The distribution
    $\pxyz(0,1,1)=\pxyz(0,0,0)=\pxyz(1,0,1)=\pxyz(1,1,0)=\frac{1}{4}$:
    If $Z=0$ Alice and Bob are perfectly correlated,
    and if $Z=1$ they are anti-correlated.
    In such a case, Alice can tell Bob her bit publicly, and because 
    an eavesdropper doesn't know
    Bob's bit, she would not be able to know the value of $\T$.
    Bob will however know $\T$ and
    can locally create a random pair of anti-correlated bits
    or correlated bits depending on the
    value of $\T$.  Thus, the distribution merging is achieved with one bit of
    public communication and no private communication.
    This reminds one of the state merging
    problem for the quantum state
    $\rho_{AB}=\frac{1}{2}(\proj{00}+\proj{11})$ whose
    purification on $R$ is the GHZ state where the merging
    is achieved with one bit
    of classical communication and no quantum communication.
    Another potential classical analog of the GHZ
    is the distribution $\pxyz(1,1,1)=\pxyz(0,0,0)=1/2$ \cite{collins-popescu},
    which has perfect correlations for all sites
    like for the GHZ state; it also has a merging cost of zero (although zero
    classical communication unlike in the quantum case).
    A distribution which has both the above features of the GHZ is
    the distribution with an equal mixture of
    $\{111,122,212,221,333,344,434,444\}$ inspired by \cite{spekkens-toy}. 
    It has perfect correlations ($1$ or $2$ on
    one site is correlated with $1$ or $2$ on the others, and likewise
    for $3$ and $4$), as well as the ability
    of one of the parties to create secret key by informing the
    other parties of her variable. Like the first GHZ like candidate,
    it also has no secret communication cost for distribution
    merging, and public communication cost of one bit,
    reminiscent of the quantum GHZ state. 
\end{enumerate}

{\bf\emph{Proof of Theorem~\ref{thm:main}.}}
We now describe the general protocol
for distribution merging. We will give two proofs of
achievability: the first is very simple and uses recycling of
the initial secret key resources. Namely, let Alice make
her transmission of Slepian-Wolf coding \cite{slepian-wolf}
secret, using a rate of $H(X|Y)$ secret bits. This gives Bob
knowledge of $XY$, which by the bi-disjointness of $P_{XYZ}$ informs
him of $Z$ [rather, the label $I$ in (\ref{eq:bi-disjoint})].
Hence he can produce a fresh sample $\widehat{X}\widehat{Y}$
of the conditional distribution $P_{XY|Z}$ -- this solves the
merging part. Now only observe that Alice and Bob are still left
with the shared $X$; from it they can extract $H(X|Z)$ secret bits
via privacy amplification \cite{privacy-BBCM}, i.e.~random hashing. By
repeatedly running this protocol, we can recover the startup
cost of providing $H(X|Y)$ secret bits, which is only later
recycled -- at least if the rate (\ref{eq:rate}) is positive. In
the appendix we show a direct proof in one step, which produces
secret key if (\ref{eq:rate}) is negative without the need to
provide some to start the process.

Now we turn to the converse, namely that this protocol is optimal.
Just as in state merging, the proof comes from looking at
monotones.  Assuming first that secret key is consumed in the
protocol, then the initial amount of secrecy that Bob has with
Alice and the reference $R$ is $H(K)+I(Y:XZ)$ where $K$ is a
random variable describing the key. By monotonicity of secrecy
under local operations and public communication this must be
greater than the final amount of secrecy he has with them; but
since he then has $\widehat{X}\widehat{Y}$, this is $I(\widehat{X}\widehat{Y}:Z) =
I(XY:Z)$. Hence $H(K) \geq I(XY:Z)-I(XZ:Y) = I(X:Z)-I(X:Y)$ as
required. If key is acquired in the protocol, then the value
$H(K)$ should be put as part of the final
amount of secrecy, and we have again $H(K)\leq I(X:Y)-I(X:Z)$.
\hfill $\Box$

The cost of distribution merging might appear quite different to
the cost of quantum state merging.  Actually this is not the case.
Since $\ket\psi_{ABR}$ is pure, we may rewrite \beq
  S(A|B)=\frac{1}{2}[ I(A:R)-I(A:B) ],
\eeq in terms of the quantum mutual information
$I(A:B):=S(A)+S(B)-S(AB)$. This looks like the cost of
distribution merging, only with a mysterious factor of $1/2$. The
factor is the same one that accounts for the fact that while one
bit of secret key has $I(A:B)=1$ and can be used in a one-time pad
protocol for one bit of secret communication, a singlet has
$I(A:B)=2$ but can teleport only one qubit.
%
For an alternative explanation, see also \cite{compl}.

\medskip
{\bf Pure and mixed state analogues.} Note that a crucial part of
the merging protocol is that once Bob knows Alice's variable, he
effectively knows $Z$ and can thus recreate the distribution (more
precisely, he knows the product distribution he shares with $R$).
Recreating the distribution would not be as easy if the total
distribution $\pxyz$ were not bi-disjoint, which further serves to
motivate our definition of bi-disjoint distributions as the
analogues of pure quantum states (although only for this
particular merging task). Nevertheless, one might wonder if we
have not overly restricted our model. Let us go back to a general
distribution $P_{XYZ}$ of Alice, Bob and the reference, and
observe that it can always be written \beq
  \label{eq:P-purification}
  P_{XYZ} = (\ID_{XY}\otimes\Lambda) \widetilde{P}_{XY\widetilde{Z}},
\eeq
with $\ID_{XY}$ the identity,
$\widetilde{P}_{XY\widetilde{Z}}$ a bi-disjoint distribution,
and a noisy channel (a stochastic map) $\Lambda:\widetilde{Z}\rightarrow Z$.
Up to relabelling of $\widetilde{Z}$ there is in fact a unique
\emph{minimal} distribution, denoted $\overline{P}_{XY\overline{Z}}$, in the
sense that every other $\widetilde{P}$ can be degraded to
$\overline{P}$ by locally applying a (deterministic) channel
$\widetilde{\Lambda}:\widetilde{Z}\rightarrow \overline{Z}$. One way of
doing this is by having $\overline{Z}$ be a record of which probability
distribution needs to be created, conditional on each $XY$.  A channel
can then act on the record $\overline{Z}$ to create the needed probability
distribution $P_{Z|XY}$. I.e. we define (cf.~\cite{wolf-wulli})
\[
  \overline{Z} = \Phi(XY) := P_{Z|XY},
\]
as an element of the probability simplex -- this means that pairs
$XY$ are labelled by the same $\overline{Z}$ (which is a deterministic
function $\Phi$ of $XY$) if and only if the
conditional distributions $P_{Z|XY}$ are the same. The channel
$\Lambda$ has the transition probabilities
$\Lambda(z|\overline{z}) = \overline{z}(z) = P_{Z=z|XY}$.
Note that $\overline{P}$ is indeed bi-disjoint.
Let us call this $\overline{P}_{XY\overline{Z}}$ the \emph{purified version}
of $P_{XYZ}$. Note the beautiful analogy to the quantum case, where
every mixed state $\rho_{ABR}$ on $ABR$ can be written
\[
  \rho_{ABR} = (\ID_{AB}\otimes\Lambda)\psi_{AB\overline{R}},
\]
with a quantum channel $\Lambda:\overline{R}\rightarrow R$ and an
essentially unique pure state $\psi_{AB\overline{R}}$ (up to local
unitaries).

\begin{theorem}
  \label{thm:mixed-main}
  For general $P_{XYZ}$, the optimal rate of distribution merging
  is that of the purified version $\overline{P}_{XY\overline{Z}}$, i.e.
  \begin{equation}
    \label{eq:purified-rate}
    I(X:\overline{Z})-I(X:Y) = H(X|Y)-H(X|\overline{Z}).
  \end{equation}
\end{theorem}
Clearly, it is achievable: we have a protocol at this rate
for $\overline{P}_{XY\overline{Z}}$, which must work for $P_{XYZ}$
as well, since the latter is obtained by locally degrading
$\overline{Z}\rightarrow Z$ which commutes with the merging
protocol acting only on Alice and Bob and makes the secrecy
condition for the final key only easier to satisfy.

To show that the rate (\ref{eq:purified-rate}) is optimal, we shall
argue that successful merging with reference $Z$ implies
that the protocol is actually successful for reference $\overline{Z}$,
at which point we can use the previous converse for ``pure''
(bi-disjoint) distributions.
Observe that Bob at the end of the protocol has to produce
samples $\widehat{X}^n\widehat{Y}^n$ such that
$P_{\widehat{X}^n\widehat{Y}^nZ^n} \approx P_{X^nY^nZ^n}$.
Assume now that it were true that with high probability (over the
joint distribution of $X^nY^nZ^n\widehat{X}^n\widehat{Y}^n$),
\begin{equation}
  \label{eq:Phi-equal}
  \widetilde{Z}^n := \Phi^n(\widehat{X}^n\widehat{Y}^n)
                       \stackrel{\text{!}}{=} \Phi^n(X^nY^n) = \overline{Z}^n.
\end{equation}
This in fact implies that merging is achieved for the
distribution $\overline{P}_{XY\overline{Z}}$:
\begin{equation*}\begin{split}
  &\bigl\| \overline{P}_{X^nY^n\overline{Z}^n}
               - \overline{P}_{\widehat{X}^n\widehat{Y}^n\overline{Z}^n} \bigr\|_1  \\
  &\phantom{==:}
   \leq \bigl\| \overline{P}_{X^nY^n\overline{Z}^n}
               - \overline{P}_{\widehat{X}^n\widehat{Y}^n\widetilde{Z}^n} \bigr\|_1 \\
  &\phantom{==:=============}
       + \bigl\| \overline{P}_{\widehat{X}^n\widehat{Y}^n\widetilde{Z}^n}
                - \overline{P}_{\widehat{X}^n\widehat{Y}^n\overline{Z}^n} \bigr\|_1 \\
  &\phantom{==:}
   \leq \| P_{X^nY^n} - P_{\widehat{X}^n\widehat{Y}^n} \|_1
       + 2\,\Pr\{ \widetilde{Z}^n \neq \overline{Z}^n\},
\end{split}\end{equation*}
and both final terms are small.
Furthermore, the secret key (possibly) distilled
at the end of the protocol has to be uncorrelated to
$\widehat{X}^n\widehat{Y}^n$,
and since this data includes knowledge of $\overline{Z}^n$, the key
will not only be secret from a reference $Z^n$ but even against $\overline{Z}^n$.

Now, unfortunately we cannot argue (\ref{eq:Phi-equal}) for a
given protocol (and insofar the situation is understood, it may
not even be generally true \cite{aram:personal});
however, we can modify the protocol
slightly -- in particular losing only a sublinear number of key
bits -- such that (\ref{eq:Phi-equal}) becomes true. We invoke a
result on so-called ``blind mixed-state compression''
\cite{koashi-mixed,koashi:undisturbed}
(see also \cite{DurVC-compr}): notice that Bob
has to output (for most $Z$) a sample of the conditional
distribution $P_{XY|Z}$, but that Alice and Bob together have
access only to one sample of that distribution, without knowing
$Z$. The central technical result in \cite{koashi-mixed} is that
every such process must preserve a lot of correlation between the
given and the produced sample, in the sense that $\Pr\bigl\{
\Phi(\widehat{X}_I\widehat{Y}_I) \neq \Phi(X_IY_I) \bigr\}$, with random
index $I$, is small. In other words, with high probability, the
string $\Phi^n(\widehat{X}^n\widehat{Y}^n)$ is within a small Hamming ball
around $\overline{Z}^n = \Phi^n(X^nY^n)$. Since Bob knows $Y^n$
already, Alice will need to send only negligible further
information about $X^n$ to Bob (invoking Slepian-Wolf another
time) so that he can determine the correct $\overline{Z}^n$ with
high probability. On the other hand, privacy amplification incurs
only a negligible loss in rate to make the final secret key
independent of this further communication (namely just its
length), and hence of $\overline{Z}^n$. Hence, we have a protocol
that effectively puts Bob in possession of $\overline{Z}$, of
which the final secret key is independent; hence he could just
output a sample from $\overline{P}_{XY|\overline{Z}}$, which would
yield a valid and asymptotically correct protocol.


The expression in Eq.~(\ref{eq:purified-rate}), when negative and
optimised over pre-processing, was previously shown to be the rate
for secret key generation~\cite{wyner75,CsiszarKorner,AhlCsi93}. Here, as
in the quantum case, we find that distribution merging provides an
interpretation of this quantity without looking at optimisations,
and for both the positive and negative case.

Note that for given
$P_{XYZ}$, if $Q_{XYZ'}=(\ID_{XY}\otimes\Lambda)P_{XYZ}$ with
$\Lambda$ sufficiently close to the identity, the two
distributions have the same purification, leading to the
conclusion that our result on distribution merging is robust under
small perturbations of the reference. Note however that a general
perturbation of $P_{XYZ}$ by an arbitrary small change in the
probability density leads to a drastic discontinuity: namely, a
generic perturbation $Q_{X'Y'Z'}$ will have trivial purification
$\overline{Z}' = X'Y'$ because all conditional distributions
$Q_{Z'|X'Y'}$ will be different. Thus, for $Q$ the merging cost
will be $H(X'|Y')$ -- essentially Slepian-Wolf coding with Bob
outputting the very $X'Y'$ of the source, so Alice and Bob's
common knowledge of $X'$ cannot be turned into secret key.
However, this is consistent with the extreme case of $P_{XYZ} =
P_{XY}\otimes P_Z$, which has merging cost $-I(X:Y)$ since Bob can
locally produce a fresh sample from $P_{XY}$, and he can extract
$I(X:Y)$ secret bits from the correlation $XY$ with Alice.


\medskip
{\bf Distribution exchange.} We now turn to finding an analogue of
quantum state exchange \cite{OW-uncommon}. In the quantum task,
not only does Alice send her state to Bob, but Bob should
additionally send his state to Alice, which is to say that the
final state is just the initial state with Alice and Bob's shares
permuted. Amazingly, this can require less resources than if only
Alice is required to send to Bob.  In general, the number of
qubits that need to be exchanged can be said to quantify the {\it
uncommon quantum information} between Alice and Bob, because this is the
part which has to sent be to their partner. We can consider the
analogy of this, where Alice and Bob must exchange distributions.
This minimal rate of secret key clearly must be non-negative,
since Alice and Bob could otherwise continue swapping their
distribution and create unlimited secret key from some given
correlation and LOPC. Note that the rate zero is indeed possible.
The distribution $P_{XYZ}(0,0,0)= P_{XYZ}(1,1,1)= P_{XYZ}(0,1,2)=
P_{XYZ}(1,0,2)=\frac{1}{4}$, for instance, has the property that
exchanging the distribution has zero exchange cost (because it is
symmetric), while the cost of Alice merging her distribution to
Bob's is $I(X:Z)=\frac{1}{2}$.

In \cite{OW-uncommon}, a lower bound for quantum state exchange
given in terms of one-way entanglement distillation between $R$
and each of the parties was proven. A similar lower bound
$K^\rightarrow(Z\>X)+K^\rightarrow(Z\>Y)$, where
$K^\rightarrow(Z\>T)$ is the distillable key (using only one-way
communication from $R$) can be proven in the context of
distribution exchange. For upper bounds, one can introduce
protocols, for example
Slepian-Wolf coding in either direction is also possible, costing
$H(X|Y)+H(Y|X)$.
A more sophisticated protocol that is sometimes better
uses results from \cite{wyner-common}:
the rate $I(X:Z)-I(X:Y)+I(XY:W)$ can be achieved (or the same
quantity with $X$ and $Y$ interchanged, whichever is smaller);
this quantity is minimized over distributions $W$ such that
$X\text{---}W\text{---}Y$ is a Markov chain. The protocol is for
Alice to merge her $X$ to Bob, which consumes $I(X:Z)-I(X:Y)$
secret bits; then Bob locally creates not $\widehat{X}\widehat{Y}|Z$ as
with merging, but rather $W|Z$ and then $W$ is essentially
communicated back to Alice -- but by \cite{wyner-common} only a
rate $I(XY:W)$ needs to be sent. Then, based on $W$, each one
creates a sample $\widehat{X}$ and $\widehat{Y}$, respectively.

An interesting aspect of quantum state exchange is that the rate
given by the sum of both parties' minimal rate of state merging
$S(A|B)+S(B|A)$ is usually not attainable (although as noted above,
one can sometimes beat it).  This is because if Alice first merges
her state with Bob, Bob will not be able to merge his state with
Alice, but must send at the full rate $S(B)$.  This is because
after Alice merges, she is left with nothing, being unable to
clone a copy of her state. This motivates us to consider the
analogue of cloning, especially since na\"{\i}vely, classical
variables can be copied. However, we need a different kind of copying
to enable Alice and Bob to merge their distributions simultaneously:
it would be for Alice to create a fresh, independent sample from
the conditional distribution $P_{X|YZ}$ of her $X$, given $Y$ and $Z$
(which are unknown to her). If she could do that, she would
be able to merge her first sample to Bob at secret key cost
$I(X:Z)-I(X:Y)$, and then he could merge his $Y$ to her second sample
(which we designed to have the same joint distribution with $YZ$),
at cost $I(Y:Z)-I(X:Y)$. Since we know that the sum
\begin{equation*}\begin{split}
  I(X:Z) &+ I(Y:Z) - 2\,I(X:Y) \\
         &\phantom{=}= H(X|Y) \!+\! H(Y|X) \!-\! H(X|Z) \!-\! H(Y|Z)
\end{split}\end{equation*}
is not in general an achievable rate, this hypothetical cloning
cannot be always possible.
Such cloning is indeed always impossible, unless the various conditional
distributions $P_{X|YZ}$ are either identical or
have disjoint support~\cite{class-no-clo}.
Note that in this case, $P_{XYZ}$ is bi-disjoint for the cut $X$-$YZ$.
A different viewpoint is that the cloning would increase the
(secret) correlation between Alice and Bob, which of course
cannot be unless they can privately communicate; this
seems to be another way of thinking about a classical analogue
of the no-cloning principle~\cite{sandu-clone}.

\medskip
{\bf Conclusion.}
In this paper, we have described a classical analogue of negative
quantum information, and we find that the similarities between
quantum information theory and privacy theory extend very far in
this analogy (at least in the present context), including no-cloning, pure and mixed states, and
GHZ-type correlations.  Quantum state merging (with reference
systems such that the overall state is pure or mixed) and state
exchange lead to similar protocols in the case of private
distributions which have many properties in common with their
quantum counterparts. 
This is part
of a body of work exploring the similarities between entanglement
and classical correlations, which, it is hoped, will stimulate
progress in both fields, for instance, on the question of the
possible existence of bound information~\cite{gisin-wolf-99}.

\bigskip
{\bf Acknowledgments.}
This work is supported by
EU grants RESQ (IST-2001-37559),
and PROSECCO (IST-2001-39227). JO additionally
acknowledges the support of a grant from the Royal Society and the Newton Trust.
AW additionally acknowledges support from the U.K. Engineering and Physical Sciences
Research Council's ``QIP IRC'' and through a University of
Bristol Research Fellowship.

\bigskip
{\bf APPENDIX -- Direct proof of Theorem \ref{thm:main}.}
For the second, direct, proof of achievability,
we will need the sampling lemma, which is proved
in \cite{wyner-common} (see also \cite{winter-common}
and \cite{Ashlwede-Winter2002}):
\begin{lemma}
  Consider a distribution $P_{UV}$ of random variables $U$ and $V$
  (with marginals $P_V$ and $P_U$), and
  $n$ independent samples $U^nV^n=U_1V_1,\ldots,U_nV_n$ from this distribution.
  Then for every $\gamma>0$ and sufficiently large $n$, there are
  $N \leq 2^{n(I(U:V)+\gamma)}$ sequences $u^{(i)}$ from $U^n$
  such that, with
  \beq
    Q := \frac{1}{N} \sum_{i=1}^N P_{V^n|U^n=u^{(i)}},
  \eeq
  \beq
    D\left( Q \| P_{V}^{\otimes n} \right) \leq 2^{-\gamma n}.
  \eeq
  Here, $D$ denotes the relative entropy.
  Furthermore, such a family of sequences is found with high probability by
  selecting them independently at random with probability distribution
  $P_{U}^{\otimes n}$.
\end{lemma}
In such a situation we say that the distribution
of $V^n$, $P_{V^n}$ is {\it covered}
by the $N$ sequences, meaning that
the distribution $P_{V}^{\otimes n}$ is approximated
with high accuracy by choosing only slightly more
than $2^{nI(U:V)}$ sequences from $U^n$.

We achieve distribution merging using a protocol extremely reminiscent of state merging.
In state merging, one adds a maximally entangled state of dimension $nS(A|B)$ bits,
and then performs a random measurement on $\rho_A$ and the pure entanglement, the result of
which is communicated to Bob.  Here, Alice and Bob add a secret key of size $H(K)$, and
the analogy of a random measurement will be a random hash (described below),
the result of which is communicated
to Bob.  In state merging, a faithful protocol has the property that $\rho_R$ is unchanged
and Bob can decode his state to $\rho_A$ after learning Alice's measurement.  Here, a successful
protocol is likewise one which allows Bob to learn $X$, while the distribution of $R$ is
unchanged if one conditions on the result of Alice's measurement.

Let us first take the case when $I(X:\R)-I(X:Y)$ is negative.
Alice and Bob previously decide on a random binning, or {\it code},
which groups Alice's $2^{nH(X)}$ sequences into $2^{nH(X|Y)}$
sets of size just under $2^{nI(X:Y)}$.
Each of these sets are numbered by $\co$ and is called the
{\it outer code}.  Within each set,
we further divide the sequences into $2^{n[I(X:Y)-I(X:\R)]}$ sets
containing just over $2^{nI(X:\R)}$ sequences.
These smaller sets are labeled by $\ci$, the {\it inner code}.
Alice then publicly broadcasts the number ${\co}$ of the outer code that her
sequence is in  (this takes $nH(X|Y)$
bits of public communication to Bob).  Now, based on
learning $\co$, Bob will know $X^n$ by the
Slepian-Wolf theorem \cite{slepian-wolf}.  We say that
he can decode Alice's sequence.  Because the
distribution $P_{XYZ}$ is bi-disjoint, and Bob knows $X^n$
and $Y^n$, he must know $\R^n$.  He
can now create the distribution $P_{\widehat{X}\widehat{Y}|\R=z}=P_{XY|\R=z}$.
He has thus succeeded in obtaining
$\widehat{X}\widehat{Y}$ such that the overall distribution is close to $P_{XYZ}$.
Furthermore, the distribution
is private --  each set (or code) in $\co$ has more than $2^{nI(X:Z)}$
elements (i.e. codewords)
[recall that  there are $2^{nI(X:Y)}$ outer codewords, and $I(X:Y)\geq I(X:Z)$].
The sampling lemma then tells us that $R$'s distribution is unchanged
i.e. $P_{Z^n|{\co}=c}\approx P^n_{Z^n}$,
which means that an eavesdropper who learns which code $\co$
Alice's sequence is in, doesn't learn anything about the sequence that $R$ has.

Next, we see that Alice and Bob gain
$n[ I(X:Y)-I(X:Z) ]$ bits of secret key.  Since Alice and Bob both know $X^n$, they both
know which inner code $\ci$ it lies in, and this they use as the key. There
are $2^{n[ I(X:Y)-I(X:Z) ]}$ of them, and each contains
just over $2^{nI(X:Z)}$ codewords in it.  Thus, from the
covering lemma, $R$'s state is independent of its value, thus she (and consequently
any eavesdropper) has arbitrarily small probability of knowing its value.

Now, in the case where $I(X:\R)-I(X:Y)$ is positive, Alice and
Bob simply use $I(X:\R)-I(X:Y)$
bits of secret key.  Since each bit of key decreases
$I(X:\R)-I(X:Y)$ by $1$, they need
this amount of key until the quantity  $I(X:\R)-I(X:Y)$
is negative, and then the preceding proof
applies.  We thus see that $I(X:Z)-I(X:Y)$ bits of key are
required to perform distribution merging,
and if it is negative, one can achieve distribution
merging, while obtaining this amount of key. \hfill $\Box$

\bibliographystyle{apsrev}

\begin{thebibliography}{31}
\bibitem[{foo({\natexlab{a}})}]{foot-1}
\bibinfo{note}{Classical distributions are the only ones for which the same
  amount of secret key is needed to create them, as can be distilled from them
  \cite{wyner-common,winter-common}, a property analogous to the fact that for
  pure states the rate of maximally entangled states needed to create them is
  equal to the rate which can be distilled from them, namely $H(p)$.}
\bibitem[{foo({\natexlab{b}})}]{foot-2}
\bibinfo{note}{We could further restrict our tripartite pure analogue to be one
  which is bi-orthogonal between any splitting of the parties into two groups,
  but such a restriction is not needed.}
\end{thebibliography}

\end{document}